# A breakthrough in Speech emotion recognition using Deep Retinal Convolution Neural Networks

Yafeng Niu[1], Dongsheng Zou[1*], Yadong Niu[2], Zhongshi He[1], Hua Tan[1]

*Abstract*—Speech emotion recognition (SER) is to study the formation and change of speaker's emotional state from the speech signal perspective, so as to make the interaction between human and computer more intelligent. SER is a challenging task that has encountered the problem of less training data and low prediction accuracy. Here we propose a data augmentation algorithm based on the imaging principle of the retina and convex lens, to acquire the different sizes of spectrogram and increase the amount of training data by changing the distance between the spectrogram and the convex lens. Meanwhile, with the help of deep learning to get the high-level features, we propose the Deep Retinal Convolution Neural Networks (DRCNNs) for SER and achieve the average accuracy over 99%. The experimental results indicate that DRCNNs outperforms the previous studies in terms of both the number of emotions and the accuracy of recognition. Predictably, our results will dramatically improve human-computer interaction.

## I. INTRODUCTION

SER is using computer to analyze the speaker's voice signal and its change process, to find their inner emotions and ideological activities, and finally to achieve a more intelligent and natural human-computer interaction (HCI), which is of great significance to develop new HCI system and to realize artificial intelligence[1]-[3].

Until now, the methods of SER can be divided into two cat- egories: the traditional machine learning method and the deep learning method.

The key to the traditional machine learning method of SER is feature selection, which is directly related to the accuracy of recognition. By far the most common feature extraction methods include the pitch frequency feature, the energy-related feature, the formant feature, the spectral feature, etc. After features extracted, the machine learning method is used to train and predict Artificial Neural Network(ANN) [4]-[7], Bayesian network model [8], Hidden Markov Model (HMM)[9]-[12], Support Vector Machine (SVM)[13], [14], Gauss Mixed Model (GMM) [15], and multi-classifier fusion [16], [17]. The primary advantage of this method is that it could train model without very large data. While the disadvantage is that it is difficult to judge the quality of the feature and may lose some key features, which will decrease the accuracy of recognition. In the meantime, it is difficult to ensure the good results can be achieved in a variety of databases.

Compared with the traditional machine learning method, the deep learning can extract the high-level features [18], [19], and it has been shown to exceed human performance in visual tasks [20], [21]. Currently, the deep learning has been applied to the SER by many researchers. Yelin Kim et al [22] proposed and evaluated a suite of Deep Belief Network(DBN) models, which can capture none linear features, and that models show improvement in emotion classification performance over baselines that do not employ deep learning. However, the accuracy is only 60% ~70%; W Zheng et al [23] proposed a DBN-HMM model, which improves the accuracy of emotion classification in comparison with the state-of-the-art methods; Q Mao et al [24] proposed learning affect-salient features for speech emotion recognition using CNN, which leads to stable and robust recognition performance in complex scenes; Z Huang et al [25] trained a semi-CNN model, which is stable and robust in complex scenes, and outperforms several well-established SER features. However, the accuracy is only 78% on SAVEE database, 84% on Emo-DB database; K Han et al [26] proposed a DNN-ELM model, which leads to 20% relative accuracy improvement compared to the HMM model; Sathit Prasomphan [27] detected the emotional by using information inside the spectrogram, then using the Neural Network to classify the emotion of EMO-DB database, and got the accuracy is up to 83.28% of five emotions; W Zheng [28] also used the spectrogram with DCNNs, and achieves about 40% accuracy on IEMOCAP database; H. M Fayek [29] provided a method to augment training data, but the accuracy is less than 61% on ENTERFACE database and SAVEE database; Jinkyu Lee [30] extracted high-level features and used recurrent neural network (RNN) to predict emotions on IEMOCAP database and got about 62% accuracy, which is higher than the DNN model; Q Jin [31] generated acoustic and lexical features to classify the emotions of the IEMOCAP database and achieved four-class emotion recognition

---

**1.** College of Computer Science, Chongqing University, Chongqing 400044, China. {dszou@cqu.edu.cn}

**2.** School of Electronics Engineering and Computer science, Peking University, Beijing 100871, China

accuracy of 69.2%; S Zhang et al [32] proposed multimodal DCNNs, which fuses the audio and visual cues in a deep model. This is an early work fusing audio and visual cues in DCNNs; George Trigeorgis et al [33] combined CNN with LSTM networks, which can auto- matically learn the best representation of the speech signal from the raw time representation.

Though previous studies have achieved some results, the accuracy of recognition remains relatively low, and it is far from the practical application. In order to address the problems of small training data and low accuracy, this paper proposes DRCNNs, which consists of two parts:

1) Data Augmentation Algorithm Based on Retinal Imaging Principle (DAARIP), using the principle of retina and convex lens imaging, we get more training data by changing the size of the spectrogram.

2) Deep Convolution Neural Networks (DCNNs) [34], which can extract high-level features from the spectrogram and make precise prediction. This novel method achieves an average accuracy over 99% on IEMOCAP, EMO-DB and SAVEE database.

## II. Deep retinal convolution neural networks

As we all know, the closer we get to the object, the bigger we see it. In other words, what we see in our retina is different because of the different distance. But it doesn't affect our recognition. Since our brains have learned high-level features of the object, we can accurately identify the same thing of different sizes.

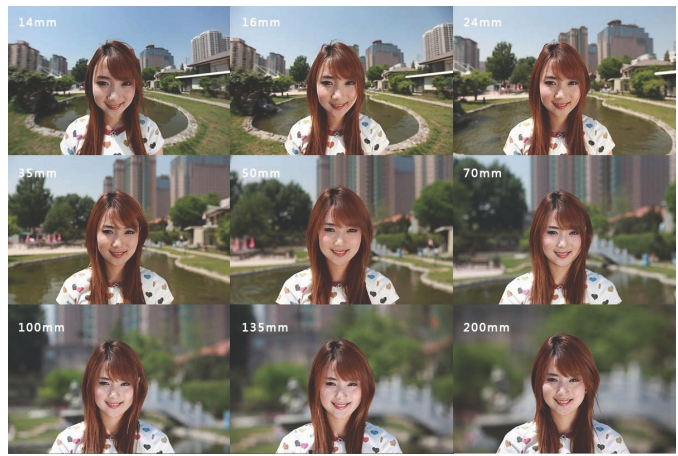

Fig. 1. Single Lens Reflex (SLR) camera is used to simulate people's retina. And it is used to simulate the same thing from different distances on the retina. The closer to the girl, the bigger the image is, and vice versa.

In Figure 1 we use the SLR camera to simulate the same thing from different distances on the retina. We can find that the closer we get to the girl, the bigger the image is, and vice versa. However, it doesn't affect our judgment. Similarly, the DCNNs is constructed from the simulation of human neurons. Two common convolution neural networks AlexNet [35] and GoogleNet have similar requirement for imputing images as 256*256 as human eyes, hence they could also make accurate judgments about the same thing in different sizes.

However, the training of deep neural network needs a large amount of data, while the data provided by current common speech emotion database is very limited. This leads to the problem that the deep neural network can't be fully trained. Referring to the imaging principle of the retinal and the convex lens, we propose DRCNNs method consisting of two parts. The working process is shown in Fig. 2.

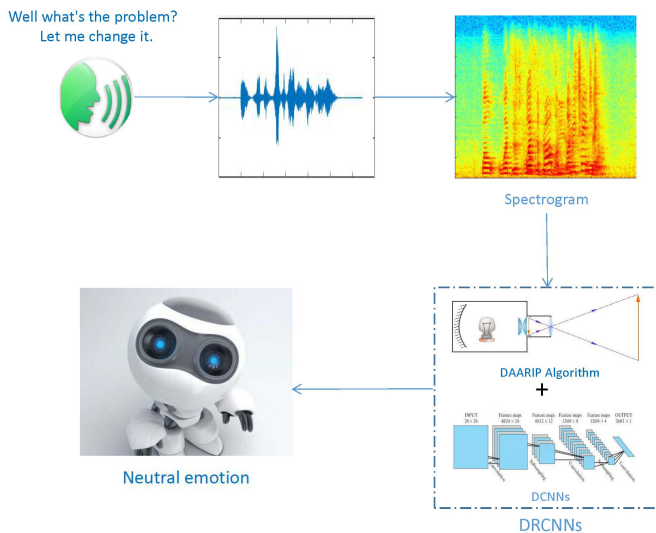

Fig. 2. The working process of DRCNNs. Firstly, people's voices are converted to spectrogram. Secondly, pass the spectrogram to DRCNNs. Finally, use DRCNNs for prediction.

1) Data Augmentation Algorithm Based on Retinal Imaging Principle (DAARIP). As shown in Table 1 and Fig. 3.

TABLE 1. PSEUDO-CODE OF DAARIP ALGORITHM

| DAARIP | |
|---|---|
| Input | Original audio data. |
| Output | Spectrograms in different size. |
| Step1 | Read audio data from file. |
| Step2 | The speech spectrogram is obtained by short time Fourier transform. (nfft = 512, window = 512, numoverlap = 384) |
| Step3 | According to the principle of retinal imaging and convex lens imaging, take $x$ point at location $L_1$ ($F<L_1<2F$) and attain $x$ images bigger than original. |
| Step4 | Take one point at $L_2$ ($L_2=2F$) and attain the same image of original size. |
| Step5 | Take $y$ point at $L_3$ ($L_3>2F$) and attain $y$ images smaller than original. |
| Step6 | Convert all images to size 256 * 256 |

2) DCNNs. We refer to the Alexnet in the experiment and cha- nge the output of the fc8 fully connected layer to the

This work was supported by the Natural Science Foundation of China (No. 61309013) and and Chongqing Basic and frontier research projects (No. CSTC2014JCYJA40042).

number of emotions we want to classify. As shown in Fig. 4. It has 5 convolution layers, 3 pooling layers and 3 fully connected layers. The processed spectrograms are the input of DCNNs. After training, the DCNNs can classify and predict the emotions accurately.

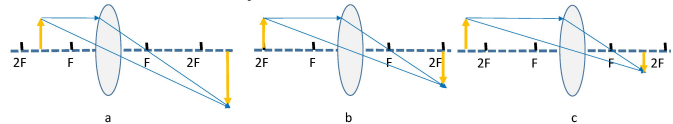

Fig. 3. Using convex lens to simulate our eyes, and take $x$ points at location $L_1$ ($F<L_1<2F$), one point at location ($L_2 = 2F$), $y$ points at location $L_3$ ($L_3 > 2F$). In this way, we can get $x+y+1$ spectrograms.

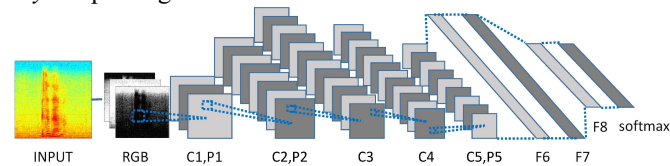

Fig. 4. The DCNNs architecture for SER using the spectrogram as input, which has 5 convolution layers (C1~C5), 3 pooling layers (P1, P2, P5) and 3 fully connected layers (F6, F7, F8).

## III. EXPERIMENTAL RESULT AND ANALYSIS

IEMOCAP database36 contains audio and label data from 10 actors, including anger, happiness, sadness, neutral, frustration, excitement, fear, surprise, disgust and other. Each utterance is labeled by 3 annotators. if their feedbacks are inconsistent with one another, the data shall be invalid. In this paper, we select 8 kinds of emotions without regard to the influence of gender.

In the first experiment, we randomly select 70% original data as training data, 15% original data as validation data, and the other 15% as testing data. After 100 epochs training on the original data, the accuracy of the validation data is achieved to 42%, and then it is over fitting. The training process is shown in Fig. 5. Next, the accuracy of the 8 types of emotions tested on the testing data is only 41.54%. From the results, it is shown that the accuracy of fear and surprise on the original data is 0%, and the accuracy of happiness is very low owing to the small training data. The accuracy of each emotion is shown in Table 2, the parameter of first experiment is shown in Table 3, and the confusion matrix on original data is shown in Table 4.

TABLE 2. EXPERIMENT ON ORIGINAL DATA

| Emotions | The original data | Accuracy |
|---|---|---|
| anger | 1103 | 38.55% |
| happiness | 595 | 2.25% |
| sadness | 1084 | 85.8% |
| neutral | 1708 | 52.73% |
| frustration | 1849 | 29.5% |
| excitement | 1041 | 30.13% |
| surprise | 107 | 0.0% |
| fear | 40 | 0.0% |
|  | 7527 | 41.54% |

TABLE 3. MAIN PARAMETER OF THE FIRST EXPERIMENT

| Parameter name | Parameter value |
|---|---|
| base_learning_rate | 0.001 |
| learning_rate_policy | fixed |
| momentum | 0.9 |
| weight_decay | 1e-05 |
| solver_type | SGD |

TABLE 4. CONFUSION MATRIX OF THE FIRST EXPERIMENT ON THE ORIGINAL TESTING DATA

|  | ang | exc | fea | fru | hap | neu | sad | sur |
|---|---|---|---|---|---|---|---|---|
| ang | 64 | 20 | 0 | 35 | 0 | 29 | 16 | 2 |
| exc | 33 | 47 | 0 | 16 | 0 | 41 | 19 | 0 |
| fea | 1 | 1 | 0 | 1 | 0 | 2 | 1 | 0 |
| fru | 25 | 25 | 0 | 82 | 0 | 84 | 62 | 0 |
| hap | 3 | 11 | 0 | 10 | 2 | 32 | 31 | 0 |
| neu | 5 | 9 | 0 | 25 | 0 | 135 | 82 | 0 |
| sad | 2 | 0 | 0 | 0 | 1 | 20 | 139 | 0 |
| sur | 2 | 0 | 0 | 2 | 0 | 3 | 9 | 0 |

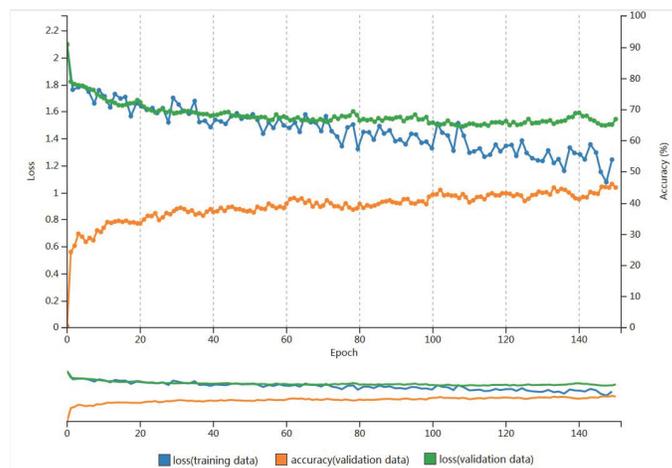

Fig. 5. The training process of DCNNs on the original data. The accuracy keeps at 42% after 100 epochs of training, then it is over fitting.

In the second experiment, we augment the original data according to the algorithm of DAARIP, and we select 70% data randomly as training data, 15% as validation data, and the other 15% as testing data. Then, training the DCNNs model on the augmented data, and the accuracy can reach 99.75% on the validation data after 20 epochs. After training, the accuracy of the 8 kinds of emotions prediction on the testing data is 99.25%. The training process is shown in Fig. 3b and the accuracy of each emotion is shown in table 5. From the results, we find that the prediction accuracy of 7 kinds of emotion attains more than 99%, and the prediction accuracy of the emotion "surprise" is 96.26%, the average accuracy on 8 emotions achieves 99.25%, which confirms that the DRCNNs can effectively solve the problems of SER. The main para- meters of second experiment is shown in

This work was supported by the Natural Science Foundation of China (No. 61309013) and and Chongqing Basic and frontier research projects (No. CSTC2014JCYJA40042).

Table 6, and the confusion matrix on the augmented data is shown in Table 7.

TABLE 5. EXPERIMENT ON AUGMENTED DATA

| Emotions | The augmented data | Accuracy |
|---|---|---|
| anger | 44213 | 99.55% |
| happiness | 29450 | 100% |
| sadness | 43360 | 99.15% |
| neutral | 70028 | 99.06% |
| frustration | 73960 | 99.17% |
| excitement | 42681 | 99.41% |
| surprise | 4815 | 96.26% |
| fear | 1560 | 100% |
|  | 310067 | 99.25% |

TABLE 6. MAIN PARAMETER OF THE SECOND EXPERIMENT

| Parameter name | Parameter value |
|---|---|
| base_learning_rate | 0.001 |
| learning_rate_policy | fixed |
| momentum | 0.9 |
| weight_decay | 1e-05 |
| solver_type | SGD |

TABLE 7. CONFUSION MATRIX OF THE SECOND EXPERIMENT ON THE AUGMENTED TESTING DATA

|  | ang | exc | fea | fru | hap | neu | sad | sur |
|---|---|---|---|---|---|---|---|---|
| ang | 6602 | 3 | 0 | 6 | 0 | 7 | 0 | 14 |
| exc | 5 | 6364 | 0 | 2 | 0 | 23 | 6 | 2 |
| fea | 0 | 0 | 234 | 0 | 0 | 0 | 0 | 0 |
| fru | 30 | 17 | 0 | 11002 | 0 | 20 | 25 | 0 |
| hap | 0 | 0 | 0 | 0 | 4462 | 0 | 0 | 0 |
| neu | 5 | 10 | 0 | 7 | 9 | 10405 | 65 | 3 |
| sad | 0 | 2 | 0 | 0 | 2 | 44 | 6449 | 7 |
| sur | 0 | 0 | 0 | 1 | 0 | 17 | 9 | 695 |

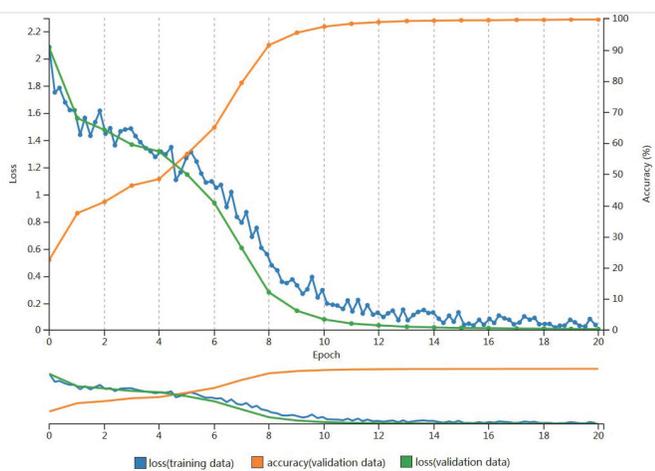

Fig. 6. The training process of DRCNNs on the augmented data. The accuracy achieves at 99.75% on the validation data after 20 epochs of training.

Compared with other recent studies, we can find that our results are better from both the number of emotions and accuracy, the detail is shown in Table 8.

TABLE 8. COMPARED WITH OTHER STUDIES

| Method | Year | Database | Emotions | Accuracy |
|---|---|---|---|---|
| Ref [31] | 2015 | IEMOCAP | 4 classes | 69.2% |
| Ref [30] | 2015 | IEMOCAP | 4 classes | 63.89% |
| Ref [28] | 2015 | IEMOCAP | 5 classes | 40.02% |
| Our Work |  | IEMOCAP | 8 classes | 99.25% |

In order to test the robustness of our proposed DRCNNs method, we do experiments on EMO-DB [37] and SAVEE database.

On the EMO-DB database, we augment the original data according to the algorithm of DAARIP, and we select 70% data randomly as training data, 15% as validation data, and the other 15% as testing data. Then, training the DCNNs model on the augmented data, and the accuracy can reach 99.9% on the validation data after 10 epochs. After training, the accuracy of the 7 kinds of emotions prediction on the testing data is 99.79%. The training process is shown in Fig. 7 and the accuracy of each emotion is shown in Table 9, the confusion matrix is shown in Table 10.

TABLE 9. EXPERIMENT ON THE AUGMENTED DATA OF EMO-DB

| Emotions | The augmented data | Accuracy |
|---|---|---|
| fear | 1320 | 100% |
| disgust | 1380 | 100% |
| happiness | 1360 | 99.51% |
| boredom | 1440 | 100% |
| neutral | 1482 | 100% |
| sadness | 1364 | 100% |
| anger | 1386 | 99.04% |
|  | 9732 | 99.79% |

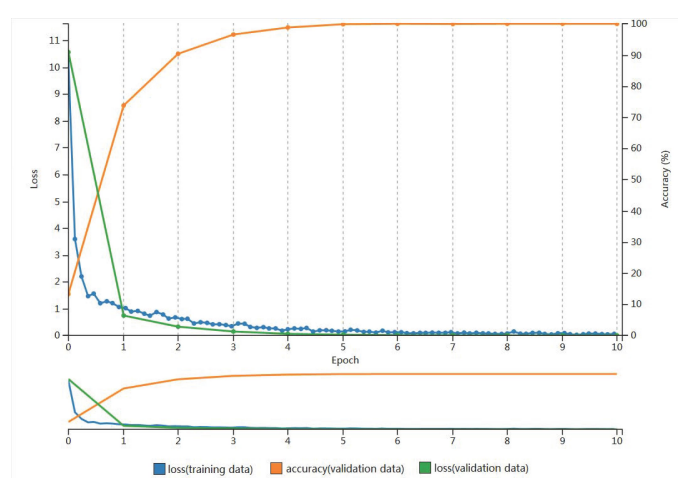

Fig. 7. The training process of DRCNNs on the augmented data of EMO-DB. The accuracy achieves at 99.9% on the validation data after 10 epochs of training.

This work was supported by the Natural Science Foundation of China (No. 61309013) and and Chongqing Basic and frontier research projects (No. CSTC2014JCYJA40042).

TABLE 10. CONFUSION MATRIX ON THE AUGMENTED TESTING DATA OF EMO-DB.

|     | fea | dis | hap | bor | neu | sad | ang |
|-----|-----|-----|-----|-----|-----|-----|-----|
| fea | 198 | 0   | 0   | 0   | 0   | 0   | 0   |
| dis | 0   | 207 | 0   | 0   | 0   | 0   | 0   |
| hap | 0   | 0   | 203 | 0   | 0   | 0   | 1   |
| bor | 0   | 0   | 0   | 216 | 0   | 0   | 0   |
| neu | 0   | 0   | 0   | 0   | 223 | 0   | 0   |
| sad | 0   | 0   | 0   | 0   | 0   | 204 | 0   |
| ang | 0   | 0   | 2   | 0   | 0   | 0   | 206 |

On the SAVEE database, we augment the original data according to the algorithm of DAARIP, and we select 70% data randomly as training data, 15% as validation data, and the other 15% as testing data. Then, training the DCNNs model on the augmented data, and the accuracy can reach 99.9% on the validation data after 5 epochs. After training, the accuracy of the 7 kinds of emotions prediction on the testing data is 99.43%. The training process is shown in Fig. 8, the accuracy of each emotion is shown in Table 11, and the confusion matrix is shown in Table 12.

TABLE 11. EXPERIMENT ON THE AUGMENTED DATA OF SAVEE

| Emotions | The augmented data | Accuracy |
|----------|--------------------|----------|
| anger    | 6368               | 100%     |
| disgust  | 6000               | 100%     |
| fear     | 6000               | 100%     |
| happy    | 6000               | 100%     |
| neutral  | 6000               | 99.78%   |
| sadness  | 6000               | 96.22%   |
| surprise | 6000               | 100%     |
|          | 42368              | 99.43%   |

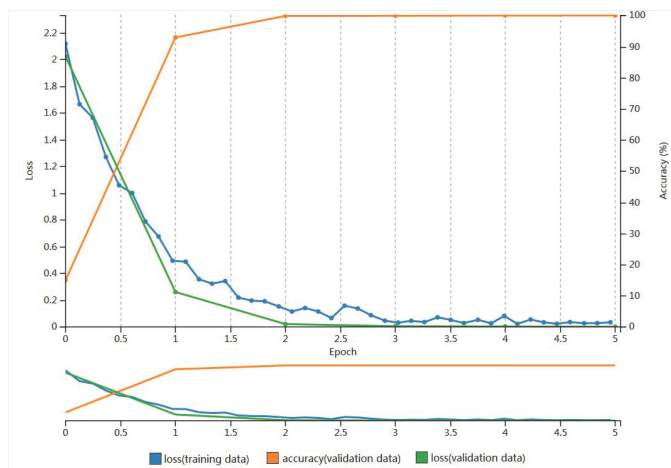

Fig. 8. The training process of DRCNNs on the augmented data of SAVEE. The accuracy achieves at 99.9% on the validation data after 5 epochs of training.

TABLE 12. CONFUSION MATRIX ON THE AUGMENTED TESTING DATA OF SAVEE.

|     | ang | dis | fea | hap | neu | sad | sur |
|-----|-----|-----|-----|-----|-----|-----|-----|
| ang | 955 | 0   | 0   | 0   | 0   | 0   | 0   |
| dis | 0   | 900 | 0   | 0   | 0   | 0   | 0   |
| fea | 0   | 0   | 900 | 0   | 0   | 0   | 0   |
| hap | 0   | 0   | 0   | 900 | 0   | 0   | 0   |
| neu | 0   | 2   | 0   | 0   | 898 | 0   | 0   |
| sad | 0   | 0   | 0   | 0   | 34  | 866 | 0   |
| sur | 0   | 0   | 0   | 0   | 0   | 0   | 900 |

These experimental results have shown the good adaptability and stability that the proposed DRCNNs method has for SER.

IV. CONCLUSION AND FUTURE WORK

SER is particularly useful for enhancing naturalness in speech based on human machine interaction. SER system has extensive applications in day-to-day life. For example, emotion analysis of telephone conversation between criminals would help criminal investigation department to detect cases. Conversation with robotic pets and humanoid partners will become more realistic and enjoyable if they are capable of understanding and expressing emotions like humans do. Automatic emotion analysis may be applied to automatic speech to speech translation systems, where speech in one language is translated into another language by the machine.

In this paper, we propose a novel method called DRCNNs, addressing the problem of small training data and low prediction accuracy in SER. The main idea of this method is two-fold. First, referring to the imaging principle of retina and convex lens, DARRIP algorithm is used to augment the original datasets, which are inputted into the DCNNs. Second, DCNNs learn high-level feature from spectrogram and classify speech emotion. Experimental results indicate that an average accuracy on three databases achieves over 99%. Obviously, the proposed method has dramatically improved the state-of –the-art in speech emotion recognition and will ultimately make major progress in HCI and artificial intelligence. We plan to extend the proposed method and evaluate its performance on multilingual speech emotion database in the near future.

This work was supported by the Natural Science Foundation of China (No. 61309013) and and Chongqing Basic and frontier research projects (No. CSTC2014JCYJA40042).

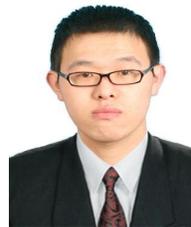

**Yafeng Niu** was born in Handan, China in 1990. He received the B.S degrees in software engineering from the University of Xinjiang. Now he is the Master candidate in the college of computer science, Chongqing university. His main research interests include machine learning, deep learning and affective computing.

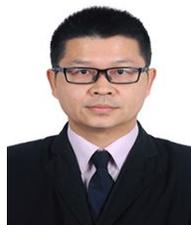

**Dongsheng Zou** received the B.S. degree, M.S. degree and Ph.D. degree in Computer Science and Technology from Chongqing University, Chongqing, China, in 1999, 2002, and 2009, respectively. He was a Postdoctoral Fellow in the College of Automation at Chongqing University from October 2009 to December 2012. He is currently an Assistant Professor in computer at Chongqing University, and a Member of China Computer Federation . His current research interests include machine learning, data mining and pattern recognition.

This work was supported by the Natural Science Foundation of China (No. 61309013) and and Chongqing Basic and frontier research projects (No. CSTC2014JCYJA40042).

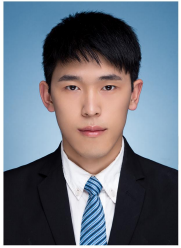

**Yadong Niu** received the B.S degrees in the School of Information and Engineering from the University of Xiamen. Now he is the Ph.D. candidate in the School of Electronics Engineering and Computer Science, Peking Universiy. His main reserach interests include machine learning, deep learning, signal processing strategies.

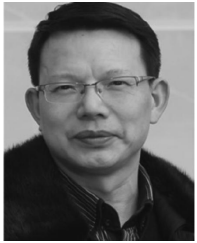

**Zhongshi He** received the B.S. degree in applied mathematics and the Ph.D. degree in computer science from Chongqing University, Chongqing, China, in 1987 and in 1996, respectively. He was a Postdoctoral Fellow in the School of Computer Science at the University of Witwatersrand in South Africa from September 1999 to August 2001.

He is currently a Full Professor, Ph.D. Supervisor, and the Vice-Dean of the College of Computer Science and Technology at the Chongqing University. He is a Member of AIPR Professional Committee of China Federation of Computer, a candidate of the 322-key talent project of Chongqing, and a Science and Technology Leader in Chongqing.

His research interests include machine learning and data mining, natural language computing, and image processing.

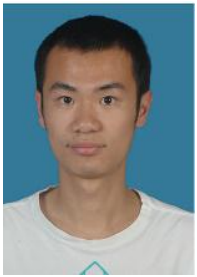

**Hua Tan** was born in Wei Fang, China in 1990. He received the B.S degrees in software engineering from Northeastern University. Now he is the Master candidate in the college of computer science, Chongqing University. His main research interests include machine learning, deep learning and affective computing.

This work was supported by the Natural Science Foundation of China (No. 61309013) and and Chongqing Basic and frontier research projects (No. CSTC2014JCYJA40042).